# Flexible Bacterial Cellulose / Permalloy nanocomposite xerogel sheets – size scalable magnetic actuator-cum-electrical conductor

V. Thiruvengadam and Satish Vitta*
Department of Metallurgical Engineering and Materials Science
Indian Institute of Technology Bombay
Mumbai 400076; India.

## Abstract

Permalloy nanoparticles containing bacterial cellulose hydrogel obtained after reduction was compressed into a xerogel flexible sheet by hot pressing at 60 ºC at different pressures. The permalloy nanoparticles with an ordered structure have a bimodal size distribution centered around 25 nm and 190 nm. The smaller nanoparticles are superparamagnetic while the larger particles are ferromagnetic at room temperature. The sheets have a room temperature magnetisation of 20 emug-1 and a coercivity of 32 Oe. The electrical conductivity of the flexible sheets increases with hot pressing pressure from 7 Scm-1 to 40 Scm-1 at room temperature.

* email: satish.vitta@iitb.ac.in

The rapid strides made by the consumer electronics industry is mainly due to the development of materials processing technologies that result in the formation of nanoscale devices. These developments in turn have led to significant decrease in the average lifetime of consumer electronic products and has resulted in the accumulation of large amounts of electronic waste.[1] Handling or treatment of this electronic waste to recover precious materials does not require any changes to be made in the very processes that create this electronic waste. Although this recovery process is very essential, it cannot be a solution in the long run. An alternative however would be to develop materials and processes that are sustainable and recyclable without causing significant damage to the environment. Towards this goal, biodegradable polymeric substrates that can host electronic circuits or nanomaterials that have electronic functions have been intensively investigated. Typical biodegradable polymeric substrate materials are – polydimethylsiloxane, low density poly-ethylene, poly-vinyl alcohol, polyimide, poly-ethylene glycol and cellulose.[2-7] Except for cellulose, all other polymers need to be synthesized while cellulose is naturally occurring and is sustainable. It is produced by both plants and bacteria. The cellulose produced by bacteria has superior physical properties because it is pure and highly crystalline compared to plant cellulose. It is environmentally friendly biopolymer with a unique structure and physical properties[8] and is amenable to functionalization. Hence, bacterial cellulose (BC) has been chosen as the host /substrate material to develop a flexible, scalable, electrically conducting and magnetically actuating composite sheet in the present work. Bimetallic alloy nanoparticles such as Pt-Cu, Pt-Pd, Pd-Ag, Ag-Cu, Pt-Co, Fe-Co, Fe-Rh, Fe-Ni have been extensively studied for their catalytic, electronic, optical and magnetic properties.[9-15] The main advantage of these bimetallic nanoparticles is that their physical properties as well as their environmental stability can be tuned/ controlled to desired values. One typical example is Fe-Ni alloys whose magnetic behavior can be continuously varied between that of Fe and Ni while the environmental stability of Fe is significantly improved by the addition of Ni. More specifically, Fe-Ni alloys in the range 70-80 at. % Ni are of extreme technological importance and are known as 'Permalloys'. These alloys exhibit an extremely soft magnetic behavior with high saturation and high permeability. The magnetostriction coefficient and magnetocrystalline anisotropy in this composition range becomes extremely small and goes to zero around 78 at. % Ni which makes them highly suitable for applications that require large permeability. They also exhibit catalytic activity [12] while retaining high thermal stability and resistance to oxidation. Since these are alloys of metals they retain high electrical conductivity. Hence, the bacterial cellulose in the present work is functionalized with $FeNi_3$ permalloy nanoparticles so that it can exhibit magnetic actuation-cum-electrical conductivity. This has been achieved by synthesizing the $FeNi_3$ nanoparticles inside bacterial cellulose matrix using an inverse chemical reduction procedure. The resulting nanocomposite hydrogels have been converted into flexible sheets by hot pressing. These sheets can be used for a variety of applications ranging from simple magnetic diaphragms to electromagnetic radiation shields and magnetic sensors.

Materials: Food grade sugar, orange peel fibers, coconut water and high purity glacial acetic acid were used to grow bacteria and synthesize bacterial cellulose. High purity, 99 % $NiCl_2.6H_2O$, $FeCl_3.6H_2O$ and $NaBH_4$ procured from Merck were used for the synthesis of permalloy nanoparticles in the bacterial cellulose matrix.

Bacterial Cellulose Synthesis: Bacterial cellulose is synthesized as per protocol reported earlier.[16, 17] The synthesis method consists of two essential steps – (i) preparation of cellulose synthesizing inoculum, and (ii) growth of cellulose using the inoculum in a static culture containing nutrients for the bacteria. The bacterial inoculum was harvested from wild culture made of orange peel fibres and sugar solution. This inoculum is allowed to grow and produce cellulose in a static nutrient medium consisting of coconut water, sugar and glacial acetic acid. The resulting cellulose is harvested and used for incorporation of $FeNi_3$ nanoparticles after a thorough cleaning with aqueous solution of NaOH followed by distilled water.

Synthesis of $FeNi_3$ – bacterial cellulose nanocomposite: A rectangular shaped bacterial cellulose sheet of size (3x1.5) $cm^2$ was first immersed in 2M concentration $NaBH_4$ aqueous solution for one hour. This process facilitates mercerizing the cellulose fibrils and also results in opening up the interfibrilar spaces. The bacterial cellulose is then transferred into the precursor aqueous solution. The precursor solution is made of 0.5 M concentrated $NiCl_2.6H_2O$ and $FeCl_3.6H_2O$ solutions in 3:1 volume ratio. The resulting black colored $FeNi_3$-BC hydrogel is washed thoroughly with distilled water to remove all unreacted solution. The hydrogel is then cut into two (1.5x1.5) $cm^2$ strips for converting into flexible xerogel sheets. The hydrogel sheet is hot pressed at 60º C under two different pressures, 9 MPa and 26 MPa to study its effect on the physical properties. These sheets are designated as FNBC9 and FNBC26 respectively.

Chacterization: The structural characterization was performed by X-ray diffraction with Cu-K$_\alpha$ radiation, field emission scanning electron microscopy (FE-SEM) and high resolution transmission electron microscopy (HR-TEM). The magnetic studies were performed using a vibrating sample magnetometer equipped with a superconducting quantum interference device detector in the temperature range 5 K to 300 K and fields ± 20 kOe. The temperature dependent electrical conductivity was studied in the range 5 K to 300 K using the physical properties measuring system.

Structure: The X-ray diffraction pattern of pristine BC obtained by hot pressing at 60º C, Figure 1 shows two sharp peaks at $2\theta \sim 14.1º$ and $22.2º$ which can be indexed to (100) and (110) planes of triclinic structure. The presence of these sharp peaks even after hot pressing indicates its robustness and highly crystalline nature. The diffraction pattern from the two nanocomposite xerogels FNBC9 and FNBC26 show a broad peak at $2\theta \sim 44.5º$ along with the peak corresponding to BC at 22.2º, Figure 1. The peak at 44.5º corresponds to (111) planes of face-centered cubic FeNi$_3$ alloy. The presence of single broad peak in the diffraction pattern clearly illustrates the nanocrystalline nature of FeNi$_3$ particles present in the BC matrix. The existence of the peak at 22.2º corresponding to BC shows that its crystallinity is not compromised by either nanoparticles inclusion or hot pressing.

The surface morphology, Figures 2(a) and 2(c), and the cross-sectional view, Figures 2(b) and 2(d) of the xerogel sheets FNBC9 and FNBC 26 respectively, as seen in the scanning electron microscope reveal several structural features. The imprints on the surface of FNBC9 xerogel sheet shows that the cellulose nanofibers that are similarly oriented become fused and laterally bonded, hydrogen bonds, to form micron sized flat ribbons due to hot pressing. These can be seen even more clearly in FNBC26 due to the effect of increased applied pressure. It should be noted here that the sheets do not have any open porosity. Both the sheets do not show the presence of FeNi$_3$ nanoparticles on the surface, clearly indicating that the nanoparticles are embedded in the cellulose matrix. This is further seen clearly in the cross-sectional micrographs which show nanoparticles in the spaces between the BC fibers. The nearly spherical nanoparticles have a log-normal distribution of sizes with average of ~ 190 nm. Such a particle size distribution is commonly observed in growth kinetics controlled nanoparticles formation.[18] An energy dispersive x-ray spectra obtained from the xerogel shows peaks corresponding to Fe, Ni, O and C indicating absence of other elemental impurities. The ratio of the elements Fe to Ni in several different areas of the nanocomposite is found to be ~ 1:3, confirming the formation of $FeNi_3$ phase. This result is in agreement with the x-ray diffraction pattern which shows the presence of only $FeNi_3$ phase together with BC in the nanocomposite sheets.

Transmission electron microscopy was performed to understand the microstructure and formation of $FeNi_3$ nanoparticles further. The transmission electron micrograph from the composite, Figure 3 shows a second size distribution of nanoparticles in the composite. The nanoparticles are mainly distributed around 25 nm and are present in the interfibrillar spaces as opposed to the larger particles which are present in voids between the fibers. These smaller sized nanoparticles however have a Gaussian distribution indicating that the growth dynamics of particles formation is different compared to that of larger nanoparticles. The larger nanoparticles formation is not constrained by space as they form in spaces between the BC fibres while formation of smaller nanoparticles is space constrained as they form in the interfibrillar spaces. The nano composite in total exhibits bimodal size distributions – the larger nanoparticles in the void spaces with a log normal distribution and the smaller nanoparticles present in the intefibrillar spaces having a Gaussian distribution. The selected area diffraction pattern from the collection of nanoparticles is shown in Figure 3(c) and shows three clear diffraction rings. An analysis of these rings indicates that they can be indexed to (110), (111) and (210) family of planes in the cubic $FeNi_3$ phase. This indicates formation of crystals with an ordered $L1_2$ cubic crystal structure instead of the disordered face centered cubic structure. Generally, $FeNi_3$ is found to exist in the disordered face centered cubic state and this disordered structure is known to transform into an ordered state at ~ 500° C. The formation of ordered $FeNi_3$ nanoparticles through solution state nucleation and growth has not been observed earlier. These results clearly show that solution phase synthesis promotes formation of near equilibrium ordered FeNi$_3$ nanoparticles. This eliminates elevated temperature heat treatment generally required for the formation of nanoparticles with an ordered crystal structure.[19]

Magnetic Studies: The magnetic behavior of FNBC xerogel sheets was studied by two different methods – (i) field dependent isothermal magnetization variation and (ii) temperature dependent iso-field magnetization variation.

The field dependent magnetization M of FNBC9 and FNBC26 in the range ± 20 kOe at 300 K, 50 K and 5 K is shown in Figures 4(a) and 4(b) respectively. Both the xerogel sheets exhibit a clear hysteresis in magnetization at all the three temperatures as seen in the insets. At room temperature and 50 K the magnetization is nearly saturated whereas at 5 K it is not saturated at a field of 20 kOe. The magnetization maximum M$_{max}$ and coercivity H$_C$ at different temperatures for FNBC9 and FNBC26 are given in Table 1. The magnetization maximum and the coercivity increase with decreasing temperature in both the cases, a typical ferromagnetic signature. The magnetization increases from ~ 19 emug$^{-1}$ to 33 emug$^{-1}$ while the coercivity increases from 32 Oe to 112 Oe in FNBC9 and 124 Oe in FNBC26 on decreasing the temperature from 300 K to 5 K. The magnetization is lower than the bulk saturation value of ~ 110 emug$^{-1}$ while the coercivity is higher by about two orders of magnitude due to finite size effects.

The temperature dependent magnetization of the xerogel sheets in the range 5 K to 300 K in the presence of varying external magnetic fields was measured both in the zero-field cooled (ZFC) and field-cooled (FC) conditions. The magnetization, Figures 5(a) and 5(b), exhibits a sharp maximum in ZFC condition, $M_{max}^{ZFC}$ at 13.6 K and 14.3 K in FNBC9 and FNBC26 respectively, called the blocking temperature T$_B$, together with a broad maximum in the range 100 K to 150 K. The magnetization between ZFC and FC conditions is irreversible, a typical behavior of magnetically interacting nanoparticles. The $M_{max}^{ZFC}$ temperature is found to vary with applied magnetic field, 200 Oe to 1500 Oe, from 14 K down to about 10 K. These results clearly show that the nanocomposites have what are termed interacting superparamagnetic nanoparticles distributed in BC matrix. The magnetization of these particles is blocked below T$_B$ while it freely rotates above this temperature similar to that observed in a classical paramagnet.

Electrical Studies: Since the objective of the work was to develop a flexible sheet which is not only magnetic but also electrically conducting, the resistivity of the sheets was studied in the temperature range 5 K to 300 K. The added advantage of studying the electrical conductivity is that it will also reveal the nature of interparticle magnetic interactions – dipolar or direct exchange that exist between the FeNi$_3$ nanoparticles. The temperature dependence of conductivity of both FNBC9 and FNBC26 is shown in Figure 6. Both the xerogel sheets exhibit a typical metallic behavior with a negative temperature coefficient. The room temperature conductivity of FNBC9 is ~ 7 Scm$^{-1}$ compared to 40 Scm$^{-1}$ for the FNBC26 sheet. The higher conductivity observed in FNBC26 is due to greater inter-particle contact achieved by hot pressing at higher pressure. These values however are still far lower than the bulk value by about 4 to 5 orders of magnitude[20] indicating that the distribution of the metallic nanoparticles is just above the percolating threshold for conductivity. Also, the nanosize of the alloy particles leads to increased scattering and hence lower conductivity. This results in increasing the temperature coefficient far above the bulk value. The conductivity at 5 K and 300 K of both the sheets is given in Table 1. The temperature dependence of conductivity of both the xerogel sheets is found to be non-linear in the complete temperature range, a typical ferromagnetic material behavior wherein the charge scattering is dominated by magnons.[21] These results also show that the interparticle magnetic interaction is not bulk like, i.e. across domains which are connected and that it is dominantly dipolar in nature.

The magnetic and electrical properties of nanoparticle composites depends on the absolute size, packing density and nature of distribution in the matrix phase. The matrix phase BC in the present nanocomposite sheets is non-magnetic and electrically insulating. Hence the FeNi$_3$ particles can be treated as isolated nanoparticles in a non-active matrix. The critical size, d$_{sp}$ for such particles above which they can sustain a stable magnetic moment against thermal demagnetization at any given temperature T is given by;[22, 23]

$$d_{sp} \approx 2 \left( 10k_B T / K_u \right)^{\frac{1}{3}} \qquad (1)$$

where $k_B$ is the Boltzmann constant and $K_u$ the effective magnetocrystalline anisotropy energy constant. The anisotropy energy constant $K_u$ depends on the chemical composition as well as crystal structure of the alloy. The chemical composition of the alloy nanoparticles determined using energy dispersive chemical analysis in the scanning electron microscope is found to be Ni – 25.8 wt.% Fe, corresponding to $Ni_3Fe$ phase and it has a face centered cubic crystal structure. The $K_u$ for this compound is reported to be ~ 2360 $Jm^{-3}$ [24] and using this value in Eq. (1) the critical size for nanoparticles that can sustain a stable magnetization at room temperature can be determined. It is found to be ~ 50 nm, implying that the FNBC sheets with a bimodal nanoparticles sizes are made of a combination of superparamagnetic and ferromagnetic nanoparticles. The nanoparticles with 25 nm average size should exhibit a superparamagnetic behavior while the larger particles with 190 nm average size are ferromagnetic in nature. The net magnetization therefore is a combination of superparamagnetic and ferromagnetic moment contributions. Since the absolute amount of $Ni_3Fe$ nanoparticles in the two xerogel sheets is identical, they should have nearly same net magnetization and it is indeed observed in Figures 5(a) and 5(b). A significant increase in the coercivity at 5 K, double the value at 50 K, is in agreement with the presence of two magnetic states in the nanocomposites. Below the superparamagnetic blocking temperature the magnetic moment of smaller nanoparticles is blocked which results in enhancing the coercivity. The irreversibility in ZFC and FC magnetizations above $T_B$ and a second broad peak in the 100 K to 150 K region is a consequence of bimodal size distributions and dipolar interparticle interactions. The presence of small superparamagnetic nanoparticles in the interfibrilar spaces in BC and formation of larger ferromagnetic nanoparticles in the voids between fibres ensures a continuous network of metallic nanoparticles which provide electrical conductivity to the nanocomposite. The conversion of hydrogel into xerogel flexible sheet at 9 MPa ensures formation of a minimal network, percolation threshold, to exhibit electrical conductivity and at 26 MPa pressure better connectivity is achieved to realise a higher conductivity.

**Conclusions**:
A flexible xerogel nanocomposite sheet with permalloy nanoparticles distributed in bacterial cellulose matrix has been successfully synthesised. The synthesis procedure is extremely simple and can be easily scaled to make large flexible sheets. The formation of nanoparticles with a bimodal size distribution using this process ensures connectivity across the BC fibres which results in a percolation path to be established on hot pressing the hydrogel into a dry xerogel flexible sheet. The electrical conductivity is found to increase with increasing the hot pressing pressure while retaining the magnetic sensitivity. These properties make it highly suitable for applications such as magnetic sensors, electromagnetic radiation shielding and so on.

**Acknowledgements**: The authors thank IIT Bombay Central Facilities for the provision of various characterization systems SEM, TEM, PPMS and MPMS.

**Table 1:**

The magnetization at 20 kOe external field $M_{max}$, coercivity $H_C$ and conductivity σ of the flexible xerogel sheets at different temperatures.

| Temperature (K) | $M_{max}$ (emu/g) | | $H_C$ (Oe) | | Conductivity (Scm$^{-1}$) | |
|---|---|---|---|---|---|---|
| | FNBC9 | FNBC26 | FNBC9 | FNBC26 | FNBC9 | FNBC26 |
| **300** | 18.7 | 18.8 | 300 | 18.7 | 6.8 | 41.1 |
| **50** | 27.8 | 27.8 | 62 | 63.4 | 7.67 | 48.6 |
| **5** | 33.3 | 32.5 | 111.5 | 124.4 | 7.67 | 49.1 |

**Figures and Figure Captions**:

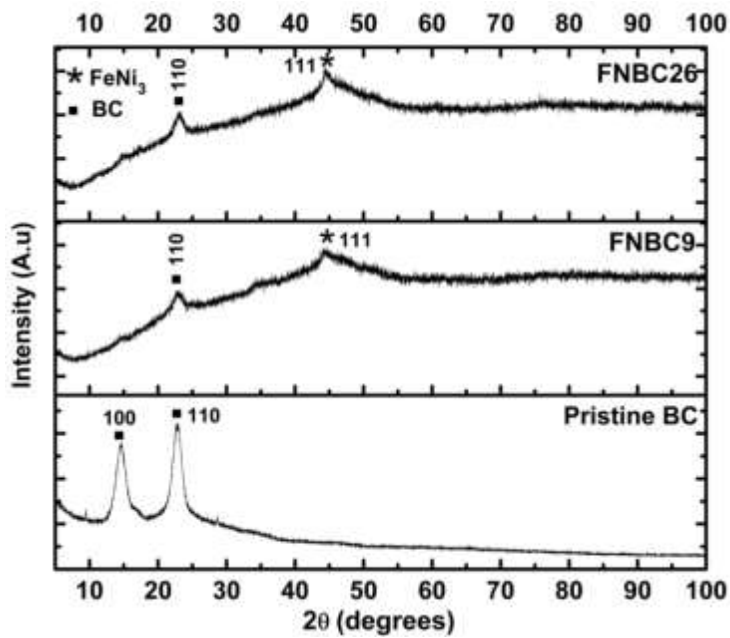

**Figure 1.** X-ray diffraction pattern of pristine bacterial cellulose (BC) shows two sharp peaks corresponding to triclinic crystal structure of cellulose while that from FNBC xerogel sheets shows a broad peak indexed to (111) planes of cubic $Ni_3Fe$ nanoparticles along with the strong peak from BC.

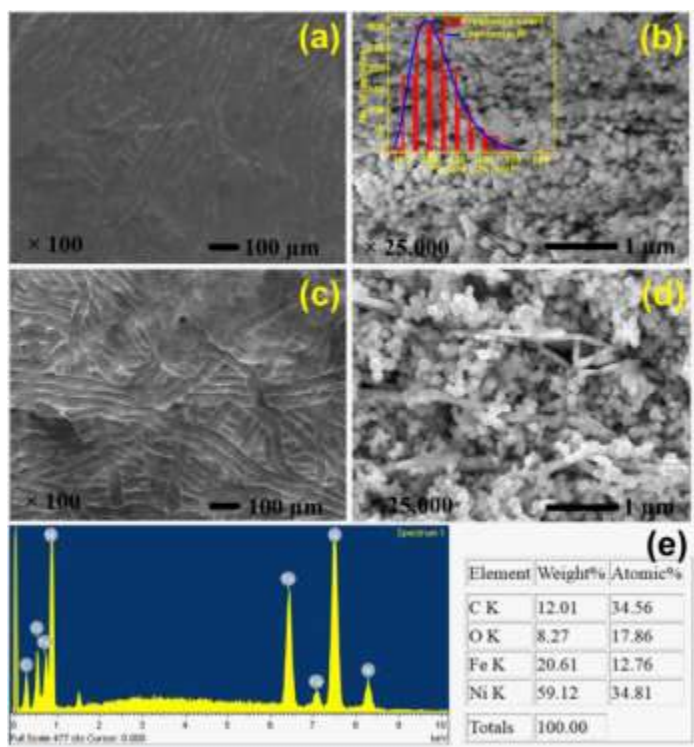

**Figure 2.** Scanning electron micrographs show the fibrous surface morphology of FNBC9 (a) and FNBC26 (c), cross-sectional view of FNBC9 (b) and FNBC26 (d). The spherical nanoparticles along with BC fibrils can be clearly seen in the cross-sectional micrographs. Energy dispersive x-ray analysis of the xerogel sheets shows peaks corresponding to Fe, Ni, C and O indicating elemental purity of the nanocomposites.

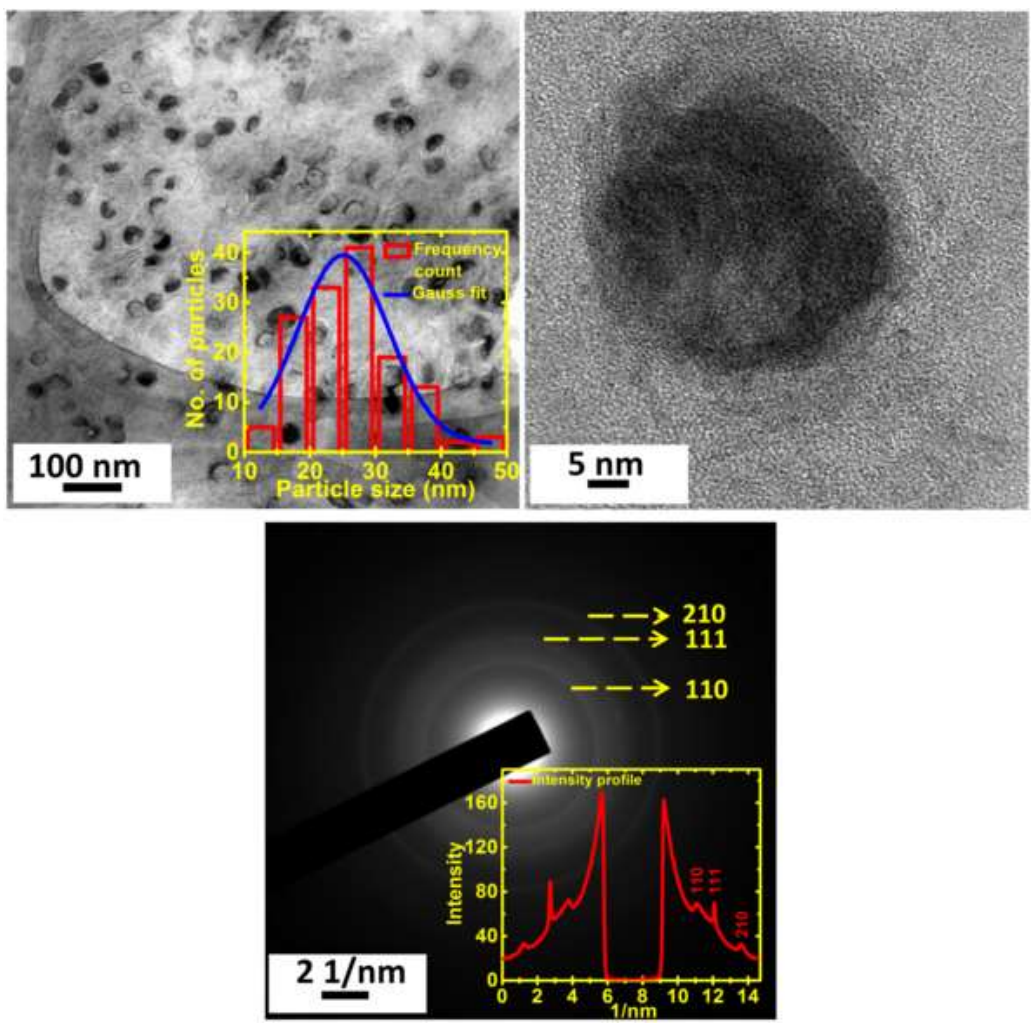

**Figure 3.** Transmission electron micrograph of the xerogel nanocomposite sheets shows clearly the presence of small $Ni_3Fe$ nanoparticles of average size 25 nm (a) and (b). The selected area diffraction pattern (c) from these nanoparticles shows three rings which can be indexed to (110), (111) and (210) planes of ordered cubic $Ni_3Fe$.

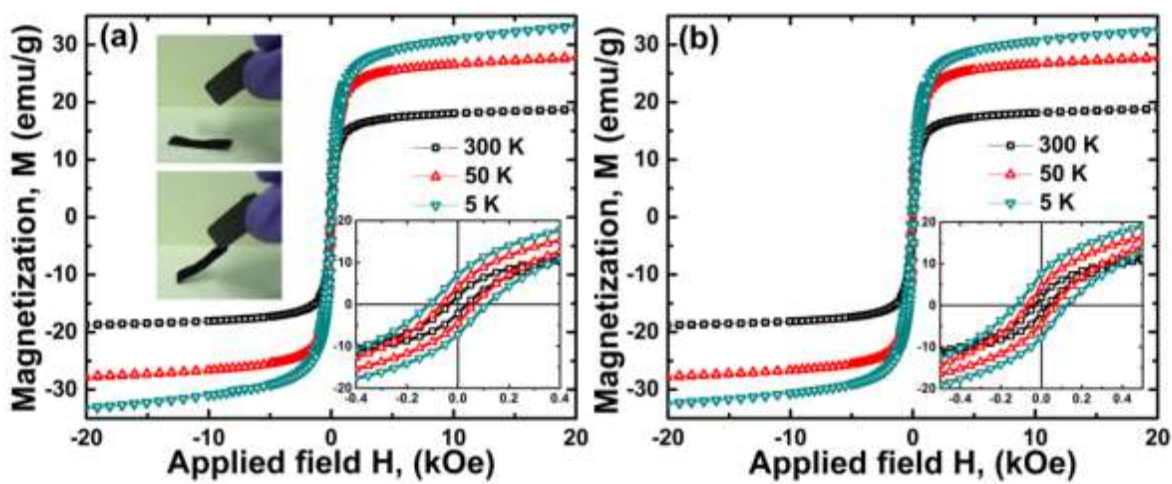

**Figure 4.** Isothermal field dependent magnetization M of FNBC9 (a) and FNBC26 (b) xerogel sheets at 5, 50 and 300 K shows hysteresis at low fields (insets). The magnetization approaches saturation at high fields. The photographs show magnetic actuation capability of the xerogel at room temperature.

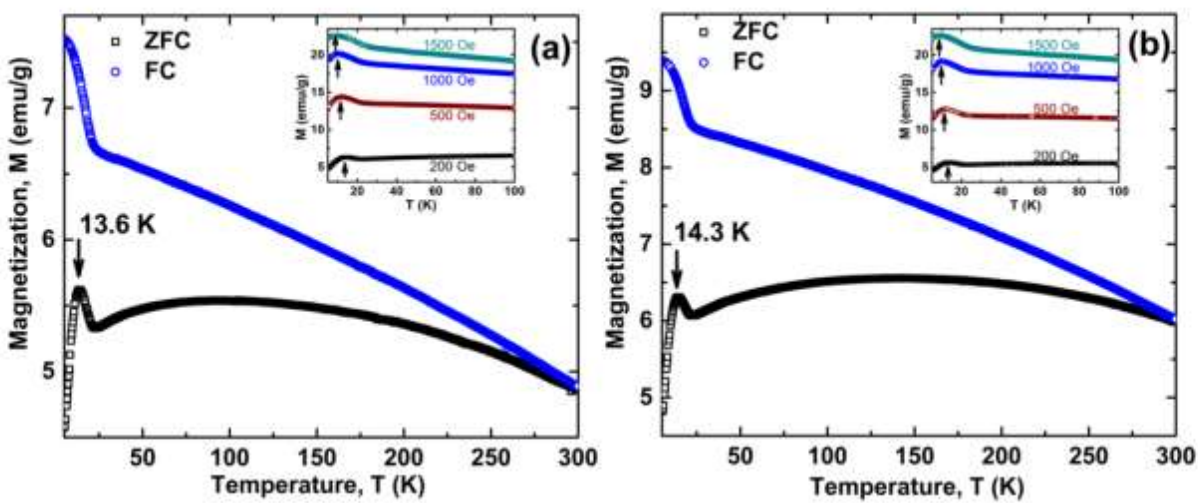

**Figure 5.** Isofield temperature dependent magnetization M of FNBC9 (a) and FNBC26 (b) in zero-field cooled (ZFC) and field-cooled (FC) condition shows irreversibility. The ZFC curve shows a low temperature peak, blocking temperature $T_B$ which shifts to lower temperatures with increasing external magnetic field (insets).

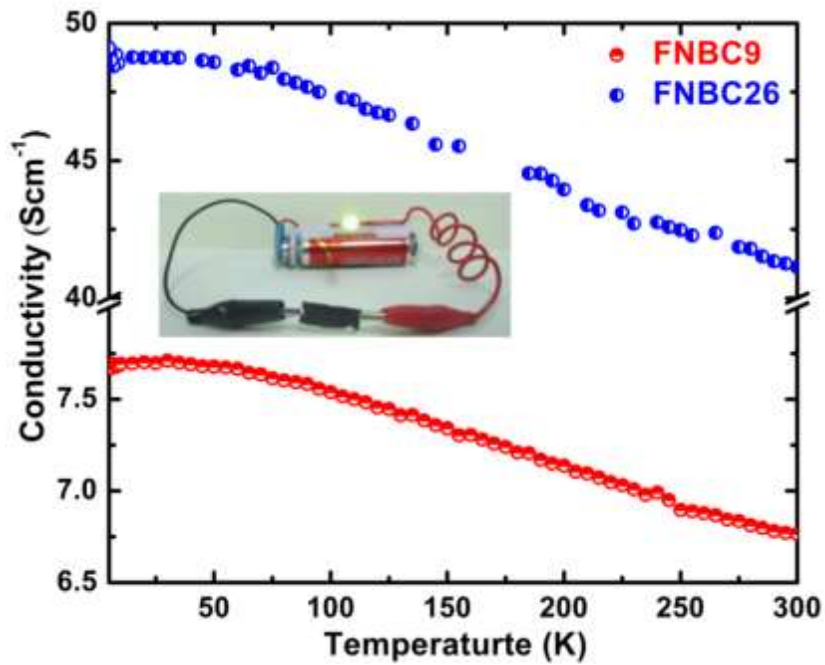

**Figure 6.** Temperature dependent d.c. electrical conductivity of FNBC9 and FNBC26 shows a metallic behavior with a negative temperature coefficient. The photograph (inset) illustrates presence of electrical conductivity at room temperature.